\documentstyle[eqsecnum,preprint,aps,12pt]{revtex}

\begin{document}
\draft

\bibliographystyle{prsty}

\title{A gauge invariant unitary theory for pion photoproduction}

\author{C.~H.~M.~van Antwerpen and I.~R.~Afnan}
\address{School of Physical Sciences,
        The Flinders University of South Australia,\\
        Bedford Park, SA 5042, Australia}
\date{\today}

\maketitle

\begin{abstract}

The Ward-Takahashi identities are central to the gauge invariance of the
photoproduction amplitude. Here we demonstrate that unitarity and in
particular the inclusion of both the $\pi N$ and $\gamma\pi N$ thresholds on
equal footing yields a photoproduction amplitude that satisfies both two-body
unitarity and the generalized Ward-Takahashi identities. The final amplitude
is a solution of a set of coupled channel integral equations for the reactions
$\pi N\rightarrow\pi N$ and $\gamma N\rightarrow\pi N$.

\end{abstract}
\pacs{ }

\newpage

\section{INTRODUCTION}\label{Sec.1}

Pion photoproduction has a long history going back to the early
50's~\cite{SB52}. In recent years the interest in this reaction has been
revived as a result of new experimental facilities which allow more careful
studies of this reaction over a much wider energy range, and the advent of
Quantum Chromodynamics (QCD) as a fundamental theory of strong interaction
which should yield the internal structure of the nucleon. In particular the
study of the reaction $p(\gamma,\pi)N$ is motivated by: (i)~The fact that this
is the simplest nuclear system  one can examine to understand the mechanism of
pion photoproduction. Furthermore the amplitude for this reaction can be used
as input into the calculation of pion photoproduction off heavier
nuclei~\cite{AA87,SO93}, with the expectation of gaining information about
nuclear structure.   (ii)~The recent interest in chiral symmetry~\cite{BH91}
and its possible violation in this reaction~\cite{MA86,BK90}, which have
raised questions regarding the role of unitarity in this
reaction~\cite{NB90,NL90}, and the use of pion photoproduction to test chiral
models of QCD, e.g.\ Chiral Perturbation Theory~\cite{BK91}.  (iii)~The new
facilities at CEBAF, and other facilities which will open the way to
examine the structure of the nucleon and the resonances observed in $\pi N$
scattering and pion photoproduction. These results could shed light on the
need to introduce quark-gluon degrees of freedom, and in particular determine
the energy at which these new degrees of freedom become more important than
the traditional meson-baryon degrees of freedom. In addition, this data could
be used to test models of QCD.

To examine the structure of resonances observed in $\pi N$ scattering using
pion photoproduction, we need to include the important thresholds for all
possible reactions at that resonance energy. To achieve this we may need to
include in our formulation those unitarity cuts, and therefore thresholds,
important for that resonance~\cite{EA89}. On the other hand the incident
photon will interact with the electromagnetic (e.m.) charge and current
distribution in the target, and to that extent it is essential to satisfy
charge conservation or $U(1)$ gauge invariance. The fact that the proton has
an internal structure, that could be described in terms of quarks at short
distances and mesons at large distances, suggests that these degrees  of
freedom should be taken into consideration in determining the electromagnetic
(e.m.) charge and current distribution. The meson degrees of freedom, which
include nucleon dressing and the distortion of the final outgoing pion wave
are part of the mechanism needed to include unitarity and therefore the proper
thresholds. Thus under ideal conditions one should satisfy both unitarity and
$U(1)$ gauge invariance in a consistent manner. In addition, the quark
dynamics should be included in the determination of the charge and current
distribution even though they do not contribute directly to any thresholds
since the quarks are confined.

Historically, early calculations of pion photoproduction were made to satisfy
unitarity by imposing the Watson Theorem~\cite{Wa54}. This basically consisted
of representing the final distortion in the $\pi N$ channel by the on-shell
pion-nucleon amplitude or $\pi N$ phase shift. In this way the cross-section
for pion photoproduction consisted of two parts, the Born amplitude which
included the interaction of the photon with the nucleon and meson currents
(see Fig.~1), and the on-shell $\pi N$ amplitude which was included to satisfy
two-body unitarity. The two ingredients were considered to be independent.
Thus for the Born amplitude, the photon coupled to a nucleon with an internal
structure represented by an on-mass-shell form factor extracted from elastic
electron-proton scattering, while the pion coupling to a finite nucleon (see
Fig.~1) involved a $\pi NN$ form factor which was constrained by the $U(1)$
gauge invariance to be the same as the e.m.\ form factor~\cite{NB90}. Here we
should note that there is no consistency between the $\pi NN$ form factor used
and the e.m.\ form factor other than the overall gauge invariance of the Born
amplitude. On the other hand, the $\pi N$ amplitude, which also has the $\pi
NN$ form factor, was determined solely by the $\pi N$ scattering data. In
other words there was no consistency between the Born amplitude for pion
photoproduction and the $\pi N$ amplitude that determined the dynamics of
pion-nucleon scattering. More recently~\cite{NB90}, extensions of this
procedure have been developed with considerable success, that extend the
Watson theorem by employing the off-shell
$\pi N$ amplitude generated by either a separable potential~\cite{NB90} or a
chiral Lagrangian~\cite{LP91} to determine the final distortion of the
outgoing pion. Even in these calculations no attempt has been made to have any
consistency between the $\pi N$ dynamics and the Born amplitude for pion
photoproduction. The present investigation is an attempt at a reformulation of
pion photoproduction that satisfies both unitarity and $U(1)$ gauge
invariance. The $\pi N$ dynamics required for unitarity will determine the
charge and current distribution. In this way we maintain consistency between
the pion dynamics and the requirement of e.m.\ gauge invariance.

To include gauge invariance and the dynamics that determine the underlying
structure of the nucleon, we need to gauge the Lagrangian for the underlying
strong interaction dynamics. Ideally this means we need to start from the QCD
Lagrangian and include the e.m.\ interaction. However at this stage we have no
practical procedure for projecting meson-baryon degrees of freedom without
resorting to models of QCD.  For example, if we assume that the Cloudy Bag
Model (CBM)~\cite{TT80,Th84} determines the dynamics of the quarks and pions,
then we should first gauge the CBM  Lagrangian~\cite{AA87} to include the
e.m.\ interaction. To substitute the quark degrees of freedom by the baryonic
degrees of freedom, we need to take the second step of implementing the
procedure adopted in the CBM by effectively integrating out the quark degrees
of freedom. This two step process of getting an effective meson-baryon
Lagrangian that includes the electromagnetic and meson interaction
consistently, is illustrated in Fig.~\ref{Fig.2}. Unfortunately, to get the
correct current for the coupling of the photon to the baryon, we need to
construct, in the CBM, a state that is an eigenstate of the total four-momentum
that can be boosted from one inertial frame to another~\cite{vA94}. This in
practice is not possible in a bag model because the bag boundary condition
cannot be simply boosted. However, this procedure could be followed in other
chiral soliton models~\cite{BG83} with some difficulty, or we can use a model
such as the NJL Lagrangian~\cite{NJ61a,NJ61b,VW91} or the Global Colour Model
(GCM)~\cite{CR85,PR87} which can in principle give a translationally invariant
state~\cite{HJ94}. The resultant Lagrangian would have gauge invariance, and
we could then proceed to implement unitarity, taking as a starting point the
Lagrangian in the space of baryons, mesons and photons. The effect of quarks
would be to have form factors for all vertices, but now we would have a
self-consistency between the different form factors in the Lagrangian. In
particular, the form factor associated with the nucleon e.m.\ current and the
form factor in the $\pi NN$ vertex would be consistent. Furthermore, this $\pi
NN$ form factor goes into the dynamics that generates the $\pi N$ interaction.
Since the quarks are confined, and therefore do not contribute to unitarity, an
alternative procedure that would satisfy gauge invariance and unitarity, see
Fig.~\ref{Fig.2}, would be to integrate the quark degrees of freedom in favour
of baryon degrees of freedom. In this case, we would have form factors
associated with the meson-baryon Lagrangian, and these form factors are related
to the quark structure of the nucleon. We then can implement
Ohta's~\cite{Oh89,Oh90} approach of minimal e.m.\ coupling to gauge the
resultant Lagrangian with form factors. The contribution to the nucleon
propagator and $\pi NN$ form factor from meson dressing will then be included
explicitly, while maintaining both unitarity and gauge invariance.

In Sec.~\ref{Sec.2} we employ Ohta's minimum electromagnetic coupling to
generate a Lagrangian that corresponds to a nucleon with internal structure
and that is gauge invariant. At this stage we will assume that this internal
structure is due to quark degrees of freedom. Furthermore, this Lagrangian in
lowest order gives a photoproduction amplitude that is gauge invariant.  This
is basically a summary of Ohta's results, and sets the stage for the next step
of including the contribution to unitarity from the pionic degrees of freedom.
This will be detailed in Sec.~\ref{Sec.3} where we derive a set of coupled
integral equations for both $\pi N$ elastic scattering and $\pi + N\leftarrow
\gamma + N$. Since the threshold energies for $\pi N$ scattering and $\gamma\pi
N$ are the same, we include both of these thresholds in our coupled channel
formulation. To establish the gauge invariance of the photoproduction
amplitude resulting from the coupled channel approach, we proceed in
Sec.~\ref{Sec.4} to make use of the Ward-Takahashi~\cite{Wa50,Ta57} identities
and the procedure proposed by Kazes~\cite{Ka59}  to derive the amplitude for
this reaction from the one pion irreducible $\pi NN$ three-point function
derived in Sec.~\ref{Sec.3}. We then show in Sec.~\ref{Sec.5} that the
amplitude resulting from the solution of the coupled channel is in fact
identical to that derived from the application of the Ward-Takahashi
identities. In this way we establish that the solution of the coupled channels
problem satisfies both unitarity and gauge invariance. Finally in
Sec.~\ref{Sec.6} we present some concluding remarks.

\section{The Gauge Invariant Lagrangian}\label{Sec.2}

Our main motivation in the present investigation is to set up a unitary and
gauge invariant formulation of pion photoproduction. Since the quarks and
gluons are confined, they do not contribute to unitarity, and  we can include
this sub-structure by introducing form factors into our Lagrangian. In other
words we can consider a Lagrangian for the system of nucleons and pions as
\begin{equation}
{\cal L} = {\cal L}_N + {\cal L}_\pi + {\cal L}_{\pi NN} \ . \label{eq:2.1}
\end{equation}
The Lagrangians for the baryon and meson with internal structure are taken to
be~\cite{Oh89}
\begin{mathletters}
\label{eq:2.2}
\begin{equation}
{\cal L}_N = \int dx'\,dx\ \bar{\psi}(x')\left[ i\gamma\cdot\partial_x r(x'-x)
-mh(x'-x) \right]\psi(x)                                     \label{eq:2.2a}
\end{equation}
\begin{equation}
{\cal L}_\pi = \int dx\,dx'\ \phi(x')\left( \partial^2_x -
\mu^2  \right)f_\pi(x'-x)\phi(x)\ .                          \label{eq:2.2b}
\end{equation}
\end{mathletters}
\noindent where the form factors $r$, $h$, and $f_\pi$ are to be determined
from the underlying quark-gluon structure, e.g.\ we could write the
corresponding propagators for the nucleon and pion as
\begin{mathletters}
\label{eq:2.3}
\begin{equation}
S_0(p) = \left[\,\not\! p - m -\Sigma(p) \,\right]^{-1}     \label{eq:2.3a}
\end{equation}
\begin{equation}
\Delta(q) = \left[\, q^2 - \mu^2 -\Pi(q^2) \,\right]^{-1}\ ,   \label{eq:2.3b}
\end{equation}
\end{mathletters}
\noindent where $\Sigma(p)$ and $\Pi(q^2)$ could be written in terms of their
quark structure as illustrated in Fig.~\ref{Fig.3}. It should be pointed out
at this stage that the QCD model used in evaluating the the diagrams in
Fig.~\ref{Fig.3} should not give rise to unitarity cuts since the quarks are
confined. The mass shifts in both the baryon and the meson propagators can
then be written in terms of the functions in the Lagrangian, e.g.\ by taking
$\Sigma(p) = \not{\! p} A(p^2) + m B(p^2)$ we can write the Fourier transform
of the functions $r$ and $h$ in terms of $A(p^2)$ and $B(p^2)$ as $r(p^2) =
1-A(p^2)$ and $h(p^2) = 1+B(p^2)$. Thus the nucleon propagator is given as
\begin{equation}
   S^{-1}_0(p)=r(p^2)\not{\! p}-m\,h(p^2) \ ,              \label{eq:2.4}
\end{equation}
The conditions that this propagator have a simple pole at the nucleon mass,
$m$, and the residue at this pole be one, impose the following conditions on
the functions $h(p^2)$ and $r(p^2)$, i.e.\
\begin{mathletters}
\label{eq:2.5}
\begin{equation}
   r(m^{2})=h(m^{2})                                       \label{eq:2.5a}
\end{equation}
and
\begin{equation}
   r(m^{2})+2m^{2}\left\{ r'(m^2) - h'(m^2) \right\} = 1   \label{eq:2.5b}
\end{equation}
with
\begin{eqnarray}
   r'(m^2)=\left.\frac{d}{dp^{2}}r(p^{2})\right|_{p^{2}=m^{2}}
                               \quad ,                     \label{eq:2.5c} \\
   h'(m^2)=\left.\frac{d}{dp^{2}}h(p^{2})\right|_{p^{2}=m^{2}}
                               \quad .                     \label{eq:2.5d}
\end{eqnarray}
\end{mathletters}
\noindent In a similar manner we can determine $f_\pi(q^2)$ in terms of
$\Pi(q^2)$ with the requirement that the pion propagator has a pole at
$q^2=\mu^2$ with unit residue. In this way the underlying quark-gluon degrees
of freedom have been included, and the gauging of these propagators will give
us the coupling of the photon to the nucleon and the pion that have internal
structure. At this stage we should point out that any pionic dressing of the
nucleon will shift the mass
$m$, and to that extent the nucleon mass in Eq.~(\ref{eq:2.5}) may need to be
considered as the bare nucleon mass. This will also apply to the pion mass,
if it is to get any further dressing.

The interaction terms in the Lagrangian in Eq.~(\ref{eq:2.1}) can be written
as
\begin{equation}
{\cal L}_{\pi NN} = \int\, dx'\,dx\,dy\,\bar{\psi}(x')\,
\Lambda_5(x',x;y)\tau_i\,\phi_i(y)\,\psi(x)  \ .           \label{eq:2.6}
\end{equation}
Here again the vertex function $\Lambda_5$ could be determined from an
underlying QCD model, e.g.\ the CBM gives the $\pi NN$ form factor in terms of
the bag radius and the wave function of the quark in the bag.

The non-locality in this Lagrangian can be turned into momentum dependence
as demonstrated by Ohta~\cite{Oh89}. This in turn allows us to gauge the
Lagrangian using the minimal substitution rule. To achieve this we follow
Ohta's~\cite{Oh89,Oh90} procedure of converting the momenta into operators,
i.e.\ $p\rightarrow \hat{p}$, and then introducing the minimal substitution
rule
\begin{equation}
\hat{p}_\mu \rightarrow \hat{p}_\mu - Q A_\mu  \ ,           \label{eq:2.7}
\end{equation}
where $Q$ is the charge operator given in terms of the nucleon isospin
operator by
\begin{equation}
   Q=e_{N}=\frac{e}{2}(1+\tau_{3})\ ,                        \label{eq:2.8}
\end{equation}
and $A_\mu$ is the photon field. Thus for the nucleon propagator we have,
using this procedure,
\begin{equation}
   S^{-1}_0(\hat{p})\rightarrow S^{-1}_0(\hat{p}-QA) \ ,     \label{eq:2.9}
\end{equation}
where after writing the propagator in terms of the form factors $r(p^2)$ and
$h(p^2)$, we have
\begin{eqnarray}
   S^{-1}_0(\hat{p}-QA)&=&\frac{1}{2}\left[r\left((\hat{p}-QA)^{2}\right)
            (\not{\! \hat{p}}-Q\not{\!\! A})
            +(\not{\! \hat{p}}-Q\not{\!\! A})
            \ r\left((\hat{p}-QA)^{2}\right)\right]  \nonumber  \\
           && \quad -mh\left((\hat{p}-QA)^{2}\right) \ .    \label{eq:2.10}
\end{eqnarray}
Clearly the presence
of the photon field in the form factors $r$ and $h$ does not allow a simple
determination of the $\gamma NN$ vertex.  As such, the photon field factors
must be extracted from $r$ and $h$.  To this end, it is assumed that there
exists a Taylor Series expansion for both $r$ and $h$ of the form
\begin{equation}
   r(\hat{p}^{2})=\sum^{\infty}_{n=0}c_{n}\hat{p}^{2n} \quad , \quad
   h(\hat{p}^{2})=\sum^{\infty}_{n=0}d_{n}\hat{p}^{2n} \quad .
                                                           \label{eq:2.11}
\end{equation}
Restricting our discussion to $r$ only and applying the minimal substitution
to equation~(\ref{eq:2.11}) gives
\begin{equation}
   r((\hat{p}-QA)^{2})=\sum^{\infty}_{n=0}c_{n}
                   \left(\hat{p}^{2}-Q(A\cdot\hat{p}+\hat{p}\cdot A)
                         +Q^{2}A^{2}\right)^{n}\ .        \label{eq:2.12}
\end{equation}
Since the charge operator $Q$, is directly proportional to the electromagnetic
coupling constant $e$, Eq.~(\ref{eq:2.12}) can be expanded in powers of $Q$.
Retaining only those terms linear in $Q$ gives,
\begin{eqnarray}
   r((\hat{p}-QA)^{2})&\approx&
     \sum^{\infty}_{n=0}c_{n}
        \biggl\{ \hat{p}^{2n}
            -Q\Bigl[
                \hat{p}^{2(n-1)}\bigl(A\cdot\hat{p}+\hat{p}\cdot A\bigr)
                +\hat{p}^{2(n-2)}\bigl(A\cdot\hat{p}+\hat{p}\cdot A\bigr)
                        \hat{p}^{2}+\cdots\nonumber \\
     &&\qquad\qquad + \bigl(A\cdot\hat{p}+\hat{p}\cdot A\bigr)\hat{p}^{2(n-1)}
              \Bigr]\biggr\}\ .                            \label{eq:2.13}
\end{eqnarray}
In the context of perturbation theory, $\hat{p}$ can be replaced by the
corresponding eigenvalue depending upon whether it acts before or after
the photon field $A$.  Since the photon gives momentum to the nucleon, as
illustrated in Fig.~\ref{Fig.4}, we can use the following substitution
\begin{equation}
   \hat{p}=\left\{\begin{array}{ll}
                     p  &\quad \mbox{if $\hat{p}$ on the right of $A$} \\
                     p' &\quad \mbox{if $\hat{p}$ on the left of $A$}
                  \end{array}
           \right. \ ,                                    \label{eq:2.14}
\end{equation}
proposed by Ohta~\cite{Oh89} to reduce equation~(\ref{eq:2.13}) to the form
\begin{equation}
   r\bigl((\hat{p}-QA)^{2}\bigr) \approx r(\hat{p}^{2})
   -Q\,\frac{r(p'^{2})-r(p^{2})}{p'^{2}-p^{2}}\, (p'+p)_\mu
        A^\mu\ .                                          \label{eq:2.15}
\end{equation}
This is the relation governing how $r$ behaves under gauging; there is a
corresponding relation for $h$.

With these results in hand, equation~(\ref{eq:2.9}), after a little
algebra involving the use of equation~(\ref{eq:2.14}), can be written to first
order in $A$ as
\begin{equation}
   S^{-1}_0(\hat{p}-QA)\approx S^{-1}_0(\hat{p})
       -\Gamma_{\mu}(k,p',p)A^{\mu}\ ,                   \label{eq:2.16}
\end{equation}
where $k$ is the photon momentum and
\begin{eqnarray}
   \Gamma_{\mu}(k,p',p) &=&\frac{Q}{2}\left(r(p'^{2})+r(p^{2})\right)
          \left[\gamma_{\mu}-\frac{\left(p'+p\right)_{\mu}}{p'^{2}-p^{2}}
             \left(\not{\! p'}-\not{\! p}\right)\right] \nonumber \\
          &&\qquad +Q\left(p'+p\right)_{\mu}
        \frac{S^{-1}_0(p')-S^{-1}_0(p)}{p'^{2}-p^{2}}\ ,  \label{eq:2.17}
\end{eqnarray}
is the resulting electromagnetic current for the nucleon. In the event that
$r(p^2)=h(p^2)=1$, i.e.\ the nucleon has no internal structure, then
$\Gamma_\mu=Q\gamma_\mu$ which is the standard Dirac current.

It is now a straight forward matter to check that $\Gamma_{\mu}(k,p',p)$,
given in Eq.~(\ref{eq:2.17}), satisfies the required Ward-Takahashi
identity~\cite{Wa50,Ta57,NK89,NK90}, i.e.\
\begin{equation}
   \left(p'-p\right)^{\mu}\Gamma_{\mu}(k,p',p)=
        Q\left\{S^{-1}_0(p')-S^{-1}_0(p)\right\}\ .       \label{eq:2.18}
\end{equation}
Thus we have the gauging of the nucleon propagator to be
\begin{eqnarray}
   S_0(\hat{p}) &\rightarrow& S_0(\hat{p} - QA) \nonumber \\
   &\approx& S_0(\hat{p}) + S_0(p')\,\Gamma_{\mu}(k,p',p)\, S_0(p)\,A^\mu
    \ ,                                                  \label{eq:2.19}
\end{eqnarray}
where perturbation theory has been used to replace $\hat{p}$ by the
corresponding  eigenvalue. In this way we have derived the nucleon current
with a form factor defined by the underlying quark model, and which satisfies
the requirement of gauge invariance to the extent that the $\gamma NN$ vertex
satisfies the Ward-Takahashi identity.

In a similar manner we can proceed to gauge the pion propagator
\begin{equation}
   \Delta^{-1}(\hat{q})=\hat{q}^{2}-m^2_{\pi}-
                        \Pi(\hat{q}^{2}) \ ,            \label{eq:2.20}
\end{equation}
where $\hat{q}$ is the pion momentum operator and $\Pi(q^2)$ contains terms
which dress the pion in terms of its quark-gluon content, e.g.\ $\bar{q}q$.
Defining $Q_{\pi}$ as the pion charge operator, we make use of the minimal
substitution prescription $q_{\mu}\rightarrow q_{\mu}-Q_{\pi}A_{\mu}$ to write
the gauged  pion propagator as
\begin{eqnarray}
   \Delta^{-1}(\hat{q})&\rightarrow&\Delta^{-1}(\hat{q}-Q_{\pi}A) \nonumber\\
   &=&(\hat{q}-Q_{\pi}A)^{2}-m_{\pi}^2
    -\Pi\bigl((\hat{q}-Q_{\pi}A)^{2}\bigr) \quad \ ,    \label{eq:2.21}
\end{eqnarray}
where we have dropped the $\mu$ index in the photon field for convenience.
Following the procedure developed above for the nucleon, we write a Taylor
series expansion for $\Pi(\hat{q}^{2})$, and expand the resultant expression
to first order in the photon field. This gives us
\begin{equation}
   \Pi((\hat{q}-Q_{\pi}A)^{2}) \approx \Pi(\hat{q}^{2})
   -Q_{\pi}(q'+q)_\mu\, \frac{\Pi(q'^{2})-\Pi(q^{2})}{q'^{2}-q^{2}}
    \, A^\mu\ ,                                        \label{eq:2.22}
\end{equation}
where
\begin{equation}
   \hat{q}=\left\{\begin{array}{ll}
                     q  &\quad \mbox{if $\hat{q}$ on the right of $A$} \\
                     q' &\quad \mbox{if $\hat{q}$ on the left of $A$}
                  \end{array}
           \right. \ ,                                \label{eq:2.23}
\end{equation}
has been used in analogy with  equation~(\ref{eq:2.14}). The gauged pion
propagator, to first order in $A$, is now given by
\begin{eqnarray}
   \Delta^{-1}(\hat{q}-Q_{\pi}A)\approx\Delta^{-1}(\hat{q})
   -\Gamma^{\pi}_{\mu}(k,q',q)A^\mu\ ,                \label{eq:2.24}
\end{eqnarray}
where $k$ is the photon momentum, and
\begin{equation}
   \Gamma^{\pi}_{\mu}(k,q',q)=Q_{\pi}\left(q'+q\right)_{\mu}
   \frac{\Delta^{-1}(q')-\Delta^{-1}(q)}{q'^{2}-q^{2}} \label{eq:2.25}
\end{equation}
is the pion electromagnetic current, which satisfies the Ward-Takahashi
identity
\begin{equation}
   \left(q'-q\right)^{\mu}\Gamma^{\pi}_{\mu}(k,q',q)= Q_{\pi}
   \left\{\Delta^{-1}(q')-\Delta^{-1}(q)\right\}\ .   \label{eq:2.26}
\end{equation}
In the limit of a structureless pion, $\Pi(q'^{2})=\Pi(q^{2})=0$,
Eq.~(\ref{eq:2.25}) reduces to the standard current for a point pion, i.e.\
\begin{equation}
   \Gamma^{\pi}_{\mu}(k,q',q)=Q_\pi(q'+q)_{\mu}\ .    \label{eq:2.27}
\end{equation}
In analogy with Eq.~(\ref{eq:2.19}) we can determine the behavior of the pion
propagator $\Delta(q)$ under gauge transformation to be
\begin{equation}
 \Delta(\hat{q})\rightarrow \Delta(\hat{q}) +
 \Delta(q')\Gamma^{\pi}_\mu(k,q',q)\Delta(q)A^\mu\ .  \label{eq:2.28}
\end{equation}

Having established the gauging of the nucleon and pion propagators, we turn
our attention to the gauging of the interaction Lagrangian, and in particular
the pion production vertex. Since we have not specified the QCD model to be
used in determining the $\pi NN$ vertex, we will consider the  most general
structure this vertex can have off-mass-shell~\cite{Ka59}, i.e.\
\begin{eqnarray}
   \Lambda^{\dagger}_5(\hat{q},\hat{p}',\hat{p})
     &=&\biggl\{i\gamma_{5}g_{1}(\hat{q}^{2},\hat{p}'^{2},\hat{p}^{2})
     +i\gamma_{5}g_{2}(\hat{q}^{2},\hat{p}'^{2},\hat{p}^{2})
        (\not{\! \hat{p}}-m) +(\not{\! \hat{p}}'-m)i\gamma_{5}
        g_{3}(\hat{q}^{2},\hat{p}'^{2},\hat{p}^{2})   \nonumber \\
 &&\quad +\ (\not{\! \hat{p}}'-m)i\gamma_{5}
        g_{4}(\hat{q}^{2},\hat{p}'^{2},\hat{p}^{2})
            (\not{\! \hat{p}}-m)\biggr\}\ ,            \label{eq:2.29}
\end{eqnarray}
where $g_{i}(\hat{q}^{2},\hat{p}'^{2},\hat{p}^{2})$ are the form factors, with
$\hat{p}$ and $\hat{p}'$ the nucleon four-momenta, and $\hat{q}$ the
pion four-momentum. We have taken the momenta to be operators since we
will be employing Ohta's approach to gauge this vertex. Since we have a
non-local vertex, we can attach the photon before or after the pion has been
emitted, and to maintain charge conservation for the overall vertex within an
isospin formalism, we need to introduce the following definitions,
\begin{mathletters}
\label{eq:2.30}
\begin{equation}
   e_{N}=\frac{e}{2}(1+\tau_{3}) \quad , \quad
   e_{L}\tau_{i}=e_{N}\tau_{i} \quad , \quad
   e_{R}\tau_{i}=\tau_{i}e_{N}                        \label{eq:2.30a}
\end{equation}
and
\begin{equation}
   (e_{L}-e_{R})\tau_{i}=
        [e_{N},\tau_{i}]=ie\epsilon_{3ij}\,\tau_{j}\ ,\label{eq:2.30b}
\end{equation}
\end{mathletters}
\noindent which enables the relevant gauge transformations to be written as
\begin{mathletters}
\label{eq:2.31}
\begin{equation}
 \hat{p}_{\mu}\rightarrow \hat{p}_{\mu}-e_{R}A_{\mu} \quad , \quad
 \hat{p}'_{\mu}\rightarrow \hat{p}'_{\mu}-e_{L}A_{\mu}\label{eq:2.31a}
\end{equation}
and
\begin{equation}
   \hat{q}_{\mu}\rightarrow \hat{q}_{\mu}-Q_{\pi}A_{\mu}
    =\hat{q}_{\mu}-(e_{R}-e_{L})A_{\mu}  \ ,          \label{eq:2.31b}
\end{equation}
\end{mathletters}
\noindent since we are considering pion emission. Applying  these gauge
transformations to the above general $\pi NN$ vertex, and proceeding with the
same steps used in gauging the propagator, i.e.\ assuming a Taylor series
expansion for the form factors
$g_i(\hat{q}^2,\hat{p}'^2,\hat{p}^2)$, we get
\begin{eqnarray}
   g_{i}(\hat{q}^{2},\hat{p}'^{2},\hat{p}^{2})
   \rightarrow g_{i}(\hat{q}^{2},\hat{p}'^{2},\hat{p}^{2})
   &-&(e_{R}-e_{L})(2q-k)_\mu\,
   \frac{g_{i}((q-k)^{2},p'^{2},p^{2})-g_{i}(q^{2},p'^{2},p^{2})}
        {(q-k)^{2}-q^{2}}\,A^\mu \nonumber \\
   &-&e_{R}(2p+k)_\mu\,
   \frac{g_{i}(q^{2},p'^{2},(p+k)^{2})-g_{i}(q^{2},p'^{2},p^{2})}
        {(p+k)^{2}-p^{2}}\,A^\mu \nonumber \\
   &-&e_{L}(2p'-k)_\mu\,
   \frac{g_{i}(q^{2},p'^{2},p^{2})-g_{i}(q^{2},(p'-k)^{2},p^{2})}
        {p'^{2}-(p'-k)^{2}}\,A^\mu\ .                 \label{eq:2.32}
\end{eqnarray}
In writing Eq~(\ref{eq:2.32}) we have made use of perturbation theory to
replace the various momenta operators by their corresponding eigenvalues,
which for photon absorption and pion emission are given by
\begin{mathletters}
\label{eq:2.33}
\begin{eqnarray}
   \hat{q}&=&\left\{\begin{array}{ll}
                q-k &\quad \mbox{if $\hat{q}$ on the right of $A$} \\
                q   &\quad \mbox{if $\hat{q}$ on the left of $A$}
                  \end{array}
           \right. \quad ,                           \label{eq:2.33a}\\
   \hat{p}&=&\left\{\begin{array}{ll}
                p   &\quad \mbox{if $\hat{p}$ on the right of $A$} \\
                p+k &\quad \mbox{if $\hat{p}$ on the left of $A$}
                  \end{array}
           \right.                                   \label{eq:2.33b}
\end{eqnarray}
\noindent and
\begin{equation}
   \hat{p}'=\left\{\begin{array}{ll}
                p'-k &\quad \mbox{if $\hat{p}'$ on the right of $A$} \\
                p'   &\quad \mbox{if $\hat{p}'$ on the left of $A$}
                  \end{array}
           \right. \ .                               \label{eq:2.33c}
\end{equation}
\end{mathletters}
\noindent Having established the procedure for gauging an individual form
factor, Eq.~(\ref{eq:2.32}), we can generate the contact term for $\pi
N\leftarrow
\gamma N$ as a result of the gauging of the $\pi NN$ form factor to be
\begin{equation}
   \Lambda^{\dagger}_5(q,p',p)\tau_i\rightarrow
   \Lambda^{\dagger}_5(q,p',p)\tau_i+\Gamma_{\mu}^{CTi}(k,q,p',p)A^{\mu}
   \ ,                                               \label{eq:2.34}
\end{equation}
where
\begin{eqnarray}
   \Gamma_{\mu}^{CTi}(k,q,p',p)
   =&&ie\epsilon_{3ij}\tau_{j}\frac{(2q-k)_{\mu}}{q^{2}-(q-k)^{2}}
   \left[\Lambda^{\dagger}_5(q,p',p)-\Lambda^{\dagger}_5(q-k,p',p)\right]
   \nonumber \\
   &&-e_{N}\tau_{i}\frac{(2p'-k)_{\mu}}{p'^{2}-(p'-k)^{2}}
   \left[\Lambda^{\dagger}_5(q,p',p)-\Lambda^{\dagger}_5(q,p'-k,p)\right]
   \nonumber \\
   &&-\tau_{i}e_{N}\frac{(2p+k)_{\mu}}{p^{2}-(p+k)^{2}}
   \left[\Lambda^{\dagger}_5(q,p',p)-\Lambda^{\dagger}_5(q,p',p+k)\right]
   \nonumber \\
   &&-i\tau_{i}e_{N}\left\{g_{2}(q^{2},p'^{2},(p+k)^{2})
        +g_{4}(q^{2},p'^{2},(p+k)^{2})(\not{\! p}'-m)\right\} \nonumber \\
             &&\hskip 4 cm \times\ \gamma_{5}\left\{\gamma_{\mu}
        +\frac{(2p+k)_{\mu}}{p^{2}-(p+k)^{2}}\not{\! k}\right\}
               \nonumber \\
   &&-ie_{N}\tau_{i}\left\{\gamma_{\mu}
        -\frac{(2p'-k)_{\mu}}{p'^{2}-(p'-k)^{2}}\not{\! k}\right\}
        \gamma_{5}\{g_{3}(q^{2},(p'-k)^{2},p^{2}) \nonumber \\
       &&\hskip 3 cm + g_{4}(q^{2},(p'-k)^{2},p^{2})(\not{\! p}-m)\}
        \ .                                            \label{eq:2.35}
\end{eqnarray}
In writing this expression for the contact term, we have restricted ourselves
to terms linear in the e.m.\ coupling, and have made use of the relations in
Eq.~(\ref{eq:2.30}) to simplify the isospin. The resultant Ward-Takahashi
identity for the contact term is then given by
\begin{eqnarray}
   k^{\mu}\Gamma_{\mu}^{CTi}(k,q,p',p)
     &=&e_{N}\tau_{i}\Lambda^{\dagger}_5(q,p'-k,p)
      -\tau_{i}e_{N}\Lambda^{\dagger}_5(q,p',p+k) \nonumber \\
     &&\hskip 2 cm -ie\epsilon_{3ij}\tau_{j}\Lambda^{\dagger}_5(q-k,p',p)\ ,
                                                       \label{eq:2.36}
\end{eqnarray}
which is in agreement with the results of Kazes~\cite{Ka59} and Naus and
Koch~\cite{NK90}.

In the present investigation these form factors will play the role of
introducing cut-offs that maintain gauge invariance and will allow us to
incorporate the pionic dressing of the nucleon without having to resort to
other renormalization procedures. From a practical point of view, the above
procedure will allow us to introduce QCD based parameters that provide
consistency between the different form factors. Although the above Lagrangian
is not the most general we can envisage, the procedures followed and the
general conclusions can be extended to more general forms for the meson-baryon
Lagrangian. In particular we could have included a $\pi N$ interaction that
under gauging would give us a term for the process $\pi N\leftrightarrow
\gamma\pi N$. The detail form of such an interaction will depend on the model
considered, e.g.\ in some chiral Lagrangians such an interaction would arise
from $\rho$ and $\sigma$ exchange~\cite{PJ91}.

\section{Unitarity}\label{Sec.3}

Having established the form of the Lagrangian, and therefore the
corresponding Hamiltonian, we now proceed to include the pionic contribution
to the currents and the amplitude for pion photoproduction. Since the pionic
dressing contributes to both the thresholds for $\pi N$ scattering and
modifies the e.m.\ currents, we need to include pionic contributions while
preserving two-body unitarity and gauge invariance. This can be achieved by
deriving coupled integral equations for pion-nucleon elastic scattering and
pion photoproduction that include the $\pi N$, $\gamma N$ as well as the
$\gamma\pi N$ thresholds. Although this latter threshold is not required for
two-body unitarity to be satisfied, the fact that it is at the same energy as
the $\pi N$ threshold suggests that we need to include it for consistency.
However, it turns out that the inclusion of the $\gamma\pi N$ threshold is
essential if we are to preserve gauge invariance at the operator level for the
pion photoproduction amplitude.

The method employed for the derivation of these coupled integral equations is
based on the classification of the Feynman diagrams that contribute to a given
amplitude, in perturbation theory, according to their irreducibility. In other
words, we take all Feynman diagrams that contribute to a given amplitude and
classify these diagrams into classes according to their irreducibility. Then
with the help of Taylors~\cite{Ta63,Ta66} last-cut lemma, we can write the
diagrams that belong to a given class, and of a given irreducibility, in terms
of amplitudes for the same process or a related process. This procedure
generates coupled integral equations for all related reactions. This method
has been used to derive unitary equations for the $NN-\pi NN$
system~\cite{TR79,Th73,AB80,AB85,AM83}, the $\pi N-\pi\pi N$
system~\cite{PA86,AP87} and the $\gamma N-\pi N$ system~\cite{AA87,Af88}.

Although Taylor's~\cite{Ta63} original classification scheme suffers from
double counting problems when applied to covariant perturbation theory, these
problems have been recently overcome~\cite{PA94,KB94} at the cost of imposing
irreducibility constraints on sub-amplitudes in channels other than the
$s$-channel, and the requirement that certain subtractions be included in the
final equations. Since this double counting does not arise for two-body
unitarity in $\pi N$ scattering and pion photoproduction~\cite{PA94}, we have
chosen to use the simplified version of the classification of diagrams that is
often used in time ordered perturbation theory, and has been applied to this
reaction previously~\cite{AA87,Af88}.

Since the last-cut lemma plays a central role in our derivation of integral
equations for pion photoproduction, we will very briefly state the basic
definition of a cut and the last-cut lemma as applied in time ordered
perturbation theory. We define a $k$-cut as an arc that separates the initial
state from the final state in a given diagram, and cuts $k$-particle lines
with at least one line being an internal line. An amplitude is $r$-particle
irreducible if all diagrams that contribute to this amplitude will not admit
any $k$-cut with $k\leq r$. With these two definitions, we can introduce the
last-cut lemma which states that for a given amplitude that is
$(r-1)$-particle irreducible, there is a unique way of obtaining an
$r$-particle cut closest to the final (initial) state for all diagrams that
contribute to the amplitude.  By virtue of this lemma, we can expose one-,
two- and three-particle intermediate states and the corresponding unitarity
cuts and in this way derive equations for the amplitude that satisfy
unitarity.  From the above statement of the lemma, it is clear that we need
to expose the $n$-particle unitarity cut before the $(n+1)$-particle
unitarity cut.

With this lemma we can now expose the $\pi N$ and $\gamma\pi N$ unitary
cuts and derive a set of integral equations governing pion-nucleon scattering
and pion photoproduction. However, before we set to derive integral equations
for $\pi N$ elastic scattering and pion photoproduction, it would be helpful
to our discussion in this section if we symbolically write our Lagrangian,
derived in Sec.~\ref{Sec.2}, as
\begin{equation}
{\cal L} = {\cal L}_N + {\cal L}_\pi + {\cal L}_\gamma + {\cal L}_{\pi NN}
           + {\cal L}_{\pi N\pi N}  +  {\cal L}_{\gamma NN}
           + {\cal L}_{\gamma\pi\pi} + {\cal L}_{\gamma\pi NN}
           + {\cal L}_{\pi N\gamma\pi N}\ .              \label{eq:3.1}
\end{equation}
The additional terms in this Lagrangian, over those in Eq.~(\ref{eq:2.1}), are
the result of gauging the Lagrangian in Eq.~(\ref{eq:2.1}). In addition we
have included terms that give rise to $\pi N\leftrightarrow\pi N$, ${\cal
L}_{\pi N\pi N}$ and $\pi N\leftrightarrow\gamma\pi N$, ${\cal L}_{\pi
N\gamma\pi N}$. These additional terms are included in the event that we need
to include $\sigma$ and $\rho$ exchange which are often needed in chiral
Lagrangians that describe $\pi N$ scattering~\cite{PJ91}. Here we will not
examine the detailed form of such a $\pi N$ interaction and the
associated gauging as the result would then be model dependent. However, we
will assume that the procedure detailed in the last section could be applied to
such an interaction maintaining gauge invariance. All terms in this Lagrangian
have form factors that are consistent with Ward-Takahashi identities as
demonstrated in Sec.~\ref{Sec.2}.

\subsection{The $\pi N$ Unitary Cut}

Restricting the analysis of the pion nucleon and pion photoproduction
amplitudes to lowest order in the electromagnetic coupling, $e$, leads to
the result that the two amplitudes can be considered separately as there
are no radiative corrections to the $\pi$-$N$ amplitude.  The anaylsis
of the $\pi$-$N$ amplitude under these conditions has been carried out
previously~\cite{AB80,AB85,AM83,PA86,AP87,AA87} and a summary of the
two-body results is presented here.  At the two-body level the $\pi$-$N$
amplitude is given by~\cite{AA87}
\begin{equation}
   t^{(0)}=t^{(1)}+
     \Lambda^{(1)\dagger}_5 S\Lambda^{(1)}_5 \quad , \label{eq:3.2}
\end{equation}
where $t$ is the $\pi-N$ amplitude, $\Lambda_5^{(1)}$ is the $N\leftarrow\pi
N$ amplitude and $S$ is the dressed nucleon propagator.  The superscript
gives the irreducibility of the amplitudes.  The one particle irreducible $\pi
N\leftarrow\pi N$ amplitude, $t^{(1)}$, satisfies the two-body equation
\begin{equation}
   t^{(1)}=t^{(2)}+t^{(2)}gt^{(1)}
          =t^{(2)}+t^{(1)}gt^{(2)} \ ,              \label{eq:3.3}
\end{equation}
with $g=S\Delta$ being the $\pi N$ propagator. Since we are restricting our
analysis to the inclusion of the $\pi N$ and $\gamma\pi N$ thresholds only, we
can choose our $\pi N$ propagator so that the nucleon propagator does not
include pionic dressing, i.e.\ $g=S_{0}\Delta$ where $S_0$ is the ``bare''
nucleon propagator defined in Sec.~\ref{Sec.2}, Eq.~(\ref{eq:2.3a}). However
the substitution $S\rightarrow S_0$, if carried through in the $\pi N$
propagator $g$, will require that $S_0$ have a simple pole at the physical
nucleon mass. The input to Eq.~(\ref{eq:3.3}) is the two particle irreducible
$\pi N$ amplitude, $t^{(2)}$. Application of the last-cut lemma to this
amplitude will expose states with two or more pions and a nucleon, i.e.\ the
$\pi\pi N$ and higher thresholds. If we are not to include these thresholds
into our final equation, then we can assume that $t^{(2)}$ is some $\pi N$
potential whose parameters can be adjusted so that the full $\pi N$ amplitude
$t^{(0)}$ reproduces the experimental phase shifts. On the other hand, if we
are to use the Lagrangian defined in Eq.~(\ref{eq:3.1}) for our input, then we
need to expose the
$\pi\pi N$ intermediate states by the application of the last-cut lemma to
$t^{(2)}$. This gives us
\begin{equation}
t^{(2)} = t^{(3)}
        + \Lambda_5^{(2)}\,S\,\Lambda_5^{(2)\dagger}\ ,\label{eq:3.4}
\end{equation}
where the second term is the cross diagram. Here again we can make the
substitution $S\rightarrow S_0$, with the condition that $S_0$ has a pole at
the physical nucleon mass, if we want to include two-body unitarity only.
Since the Lagrangian in Eq.~(\ref{eq:3.1}) does not include any term for two
pion production or absorption, $\pi\pi N\leftrightarrow N$, the three particle
irreducible amplitude $t^{(3)}$ is just the $\pi N$ potential resulting from
${\cal L}_{\pi N\pi N}$ and could include $\sigma$ and $\rho$ exchange. If we
were to examine the full $\pi\pi N$ part of the Hilbert space, then the $\pi
N\rightarrow N$ amplitude in Eq.~(\ref{eq:3.4}) gets dressed~\cite{AP87} to
the extent that $\Lambda_5^{(2)}\rightarrow\Lambda_5^{(1)}$, and the $\pi
N\rightarrow N$  amplitude in the crossed digram becomes the same as that in
Eq.~(\ref{eq:3.2}) for the pole diagram.

We now turn to the one particle irreducible $\pi N\rightarrow N$ amplitude,
$\Lambda_5^{(1)}$. Using the last-cut lemma, we can write.
\begin{equation}
 \Lambda^{(1)}_5=\Lambda^{(2)}_5+\Lambda^{(2)}_5 g\,t^{(1)}
        =\Lambda^{(2)}_5+\Lambda^{(1)}_5 g\,t^{(2)}\ .  \label{eq:3.5}
\end{equation}
Here again since the Lagrangian under consideration has no component that
gives rise to two pion production or absorption, i.e.\ $N\leftrightarrow \pi\pi
N$, then the two particle irreducible $\pi N\leftarrow N$ amplitude
$\Lambda_5^{(2)}$ has no intermediate states and is just the $\pi NN$ vertex
in the Lagrangian, i.e.\ $\Lambda_5^{(2)} = \Lambda_5$.

The dressed nucleon propagator in Eq.~(\ref{eq:3.2}) is given by
\begin{equation}
   S^{-1}=S_{0}^{-1}-\Sigma^{(1)}\ ,                   \label{eq:3.6}
\end{equation}
where, with the help of the last-cut lemma, we can write the mass shift due
to pionic dressing as
\begin{equation}
 \Sigma^{(1)}=\Sigma^{(2)}
               +\Lambda^{(1)}_5g\,\Lambda^{(2)\dagger}_5
             =\Sigma^{(2)}
               +\Lambda^{(2)}_5g\,\Lambda^{(1)\dagger}_5\ .\label{eq:3.7}
\end{equation}
Since the Lagrangian has no term for $N\leftrightarrow\pi\pi N$ transition,
then $\Sigma^{(2)}=0$, while the ``bare'' nucleon propagator  $S_{0}$ in
Eq.~(\ref{eq:3.6}) is given by
\begin{equation}
   S_{0}^{-1}=\gamma_{\mu}p^{\mu} - m_{N}^{(0)}        \label{eq:3.8}
\end{equation}
where $m^{(0)}_{N}$ is the bare nucleon mass, and includes the quark-gluon
contribution to the mass. From this point on we assume this `bare' mass has no
momentum dependence as the quark-gluon contribution does not introduce any
unitarity thresholds and the mass shift due to quark-gluon structure has no
threshold. In this case the dressed propagator $S$ should have a pole at the
physical nucleon mass.

An integral equation for the $\pi$-$N$ amplitude, $t^{(0)}$, can now be
derived~\cite{AP87}. To achieve this, we first substitute
equations~(\ref{eq:3.3}) and (\ref{eq:3.5}) into (\ref{eq:3.2}) resulting in
\begin{equation}
 t^{(0)}=t^{(2)}+\Lambda^{(2)\dagger}_5 S\Lambda^{(1)}_5
         + t^{(2)}\,g\,\left(t^{(1)}
         + \Lambda^{(1)\dagger}_5 S\Lambda^{(1)}_5\right) \ .\label{eq:3.9}
\end{equation}
We now make use of the fact that
\begin{equation}
  S\Lambda_5^{(1)}=S_{0}\Lambda_5^{(0)}
       =S_{0}\Lambda_5^{(2)}\left(1+gt^{(0)}\right)\ ,   \label{eq:3.10}
\end{equation}
and noting the definition of $t^{(0)}$, obtain
\begin{equation}
   t^{(0)}=v\,\left(1+gt^{(0)}\right)\ .                 \label{eq:3.11}
\end{equation}
Here the Born amplitude for $\pi N$ scattering $v$ is given by
\begin{equation}
   v=t^{(2)}
    +\Lambda^{(2)\dagger}_5\, S_{0}\,\Lambda^{(2)}_5     \label{eq:3.12}
\end{equation}
where for the present Lagrangian
\begin{equation}
t^{(2)} = t^{(3)} + \Lambda_5^{(2)}\,S_0\,\Lambda_5^{(2)\dagger}\quad
\mbox{and}\quad \Lambda_5^{(2)} = \Lambda_5\ ,           \label{eq:3.13}
\end{equation}
where $\Lambda_5$ is given in Eq.~(\ref{eq:2.29}) in terms of form factor
that could in principle be extracted from a QCD model and $t^{(3)}$ could be
any $\pi N$ potential we may need to introduce. This Born amplitude, or $\pi
N$ potential is illustrated in Fig.~\ref{Fig.5}.

We now turn to the amplitude for single pion photoproduction, $\pi +
N\leftarrow \gamma + N$. This has been considered in detail previously by
Araki and Afnan (AA)~\cite{AA87}. For the sake of completeness we present
here a summary of their results. Applying the last-cut lemma to the pion
photoproduction amplitude, $M^{(0)}$, we first expose the $\pi N$ unitary cut.
Here we will restrict the analysis to first order in the electromagnetic
coupling $e$. We now can divide the diagrams that contribute to this
amplitude into two classes: The diagrams that are one-particle irreducible we
sum to get the one-particle irreducible amplitude $M^{(1)}$.  The rest of the
diagrams are one particle reducible, and when summed give the nucleon pole
contribution to the full amplitude. The resultant decomposition for the full
amplitude is then given by
\begin{equation}
   M^{(0)}=M^{(1)}
          +\Lambda^{(1)\dagger}_5\, S\, \Gamma^{(1)}\ ,\label{eq:3.14}
\end{equation}
where $\Gamma^{(1)}$ is the one-particle irreducible  $N\leftarrow\gamma N$
amplitude.  Following the procedure presented above for the $\pi N$ amplitude,
we employ the last-cut lemma to write an integral equation for the
one-particle irreducible photoproduction amplitude, $M^{(1)}$. This is given
by
\begin{equation}
   M^{(1)}=M^{(2)}+t^{(2)}\,g\, M^{(1)}
          =M^{(2)}+t^{(1)}\,g\, M^{(2)}\ ,             \label{eq:3.15}
\end{equation}
where we have retained only those terms linear in $e$.  Finally,
applying the last-cut lemma to expose the $\pi N$ unitarity cut in the
$N\leftarrow\gamma N$ amplitude, $\Gamma^{(1)}$, we get
\begin{equation}
   \Gamma^{(1)}=\Gamma^{(2)}+\Lambda^{(1)}_5 gM^{(2)}
               =\Gamma^{(2)}+\Lambda^{(2)}_5 gM^{(1)}\ .\label{eq:3.16}
\end{equation}
In this way we have exposed all two body unitarity cuts in the
photoproduction amplitude.

We now write this photoproduction amplitude, $M^{(0)}$, as the solution to a
two-body integral equation, and in this way define the Born amplitude for the
reaction $N\pi\leftarrow N\gamma$. This involves similar steps to those used
to derive the equivalent equation for $\pi N$ elastic scattering, i.e.\
Eqs~(\ref{eq:3.9}) to (\ref{eq:3.12}). This results in the integral equation
for $M^{(0)}$ being
\begin{equation}
   M^{(0)}=\tilde{v}  +v\,g\,M^{(0)}\ ,               \label{eq:3.17}
\end{equation}
with the Born amplitude for pion photoproduction given by
\begin{equation}
 \tilde{v}=M^{(2)}
          +\Lambda^{(2)\dagger}_5\,S_{0}\,\Gamma^{(2)} \label{eq:3.18}
\end{equation}
where we have made use of the fact that
\begin{equation}
   S\,\Gamma^{(1)}=S_{0}\,\Gamma^{(0)}
                  =S_{0}\,\left(\Gamma^{(2)}
                +\Lambda^{(2)}_5\,g\,M^{(0)}\right)\ . \label{eq:3.19}
\end{equation}
At this stage it is premature to link the photoproduction Born amplitude
$\tilde{v}$ to the underlying Lagrangian since the last-cut lemma can be
further applied to both $\Gamma^{(2)}$ and $M^{(2)}$ to reveal the contribution
of the $\gamma\pi N$ cut. As we will observe, when we expose the $\gamma\pi N$
unitarity cut, $\tilde{v}$ is not the usual Born amplitude for pion
photoproduction. It will include additional terms that give rise to the
dressing of some of the vertices.

\subsection{The $\gamma\pi N$ Cut}

To examine the diagrams that contribute to $\tilde{v}$ and in this way
determine their physical importance, the last-cut lemma is applied to  both
$\Gamma^{(2)}$ and $M^{(2)}$ to reveal the $\gamma \pi N$ unitarity cut.  The
$\gamma \pi N$ branch point occurs at the same energy as the $\pi N$ branch
point but at a lower energy than the start of the $\pi\pi N$ cut as
illustrated in Fig.~\ref{Fig.6}.  As a result, in this approach, the
$\gamma\pi N$ cut will be treated on an equal footing with the $\pi N$ cut but
differently from the $\pi \pi N$ cut, which has been truncated out of our
analysis since we are considering only two-body unitarity as far as the pion
dynamics is concerned. In the approach of AA~\cite{AA87}, the $\pi\pi N$ and
$\gamma\pi N$ cuts were exposed simultaneously since their truncation was
carried out with respect to the number of particles present in intermediate
states in contrast with the present approach of truncation on the basis of the
position of the threshold in the energy plane.

To include the $\gamma\pi N$ threshold in our pion photoproduction amplitude,
we first consider the diagrams that contribute to the two-particle irreducible
$N\leftarrow\gamma N$ amplitude, $\Gamma^{(2)}$. These diagrams can be divided
into two classes: Those that don't have $\gamma\pi N$ intermediate states, the
sum of which we denote by $\Gamma^{(3)}$. The second class of diagrams are
those that have $\gamma\pi N$ intermediate states. To these  we apply the
last-cut lemma and thus expose the corresponding threshold. We denote the sum
of the diagrams in this class by $\left\{\Gamma_\pi^{(3)} \,\tilde{G}^{(2)}\,
\tilde{F}_2^{(2)\dagger}\right\}_c$, where the subscript $_c$ denotes that we
include only the connected diagrams in the sum. Here, the sub-amplitudes
$\Gamma_\pi^{(3)}$ for $N\leftarrow\gamma\pi N$ and $\tilde{F}_2^{(2)\dagger}$
for $\gamma\pi N\leftarrow\gamma N$ do not have to be the sum of connected
diagrams provided diagrams that contribute to $N\leftarrow\gamma N$ amplitude
$\Gamma^{(2)}$ are connected. Finally, $\tilde{G}^{(2)}$ is the the product of
the propagators for the nucleon, pion and photon. Since we have a pion in
$\tilde{G}^{(2)}$, the nucleon propagator in $\tilde{G}^{(2)}$ need not
include any pionic dressing, but the mass in this propagator needs to be the
physical nucleon mass. Thus exposing the
$\gamma\pi N$ threshold in the $N\leftarrow\gamma N$ amplitude allows us to
write this amplitude as
\begin{equation}
   \Gamma^{(2)}=\Gamma^{(3)}+\left\{\Gamma_\pi^{(3)}\,\tilde{G}^{(2)}\,
             \tilde{F}_{2}^{(2)\dagger}\right\}_{c} \ .\label{eq:3.20}
\end{equation}
In the absence of a term in the Lagrangian for the $N\leftrightarrow\pi\pi N$
transition, the amplitude $\Gamma^{(3)}$ is nothing more than the nucleon
e.m.\ current as given in Eq.~(\ref{eq:2.17}), i.e.\ $\Gamma^{(3)} = \Gamma$.
In a similar manner, the absence of a two pion production or absorption term in
the Lagrangian allows us to take the  $N\leftarrow\gamma\pi N$ amplitude,
$\Gamma_\pi^{(3)}$ to be basically that resulting from the gauging of the $\pi
NN$ vertex, and given in Eq.~(\ref{eq:2.35}), i.e.\ $\Gamma_\pi^{(3)} =
\Gamma_\pi$. Finally, taking into consideration the fact that we are
restricting all our results to the inclusion of amplitudes in lowest order in
the e.m.\ coupling, the amplitude $\tilde{F}_2^{(2)\dagger}$ for the process
$\gamma\pi N\leftarrow\gamma N$, has to be disconnected and of the form
\begin{equation}
 \tilde{F}_2^{(2)\dagger}=D^{-1}\,\Lambda^{(1)\dagger}_5\ ,\label{eq:3.21}
\end{equation}
where $D$ is the photon propagator. This allows us to write the two-particle
irreducible $N\leftarrow\gamma N$ amplitude as
\begin{equation}
 \Gamma^{(2)}=\Gamma^{(3)}
             +\Gamma_\pi^{(3)}\,g\,\Lambda^{(1)\dagger}_5\ .\label{eq:3.22}
\end{equation}
where for the present Lagrangian we have
\begin{equation}
\Gamma^{(3)} = \langle N|-{\cal L}_{\gamma NN}|\gamma N\rangle
             = \Gamma                                    \label{eq:3.23}
\end{equation}
and
\begin{equation}
\Gamma_\pi^{(3)} = \langle N|-{\cal L}_{\gamma\pi NN}|\gamma\pi N\rangle
                 = \Gamma_\pi\ ,                        \label{eq:3.24}
\end{equation}
with $\Gamma$ and $\Gamma_\pi$ resulting from the gauging ${\cal L}_N$ and
${\cal L}_{\pi NN}$ in our basic meson-baryon Lagrangian respectively.

We now turn to the two-particle irreducible  $\pi N\leftarrow \gamma N$
amplitude, $M^{(2)}$. Here again we divide all diagrams that contribute to
this amplitude into two classes: Those with $\gamma\pi N$ intermediate states.
These diagrams are summed, with the help of the last-cut lemma, to produce the
amplitude in which the $\gamma\pi N$ cut is exposed. The sum of all the
diagrams in this class is given by $\left\{\tilde{F}_3^{(2)}\,
\tilde{G}^{(2)}\,\tilde{F}_2^{(3)\dagger}\right\}_c$, where
$\tilde{F}_3^{(2)}$ is the two-particle irreducible $\pi N\leftarrow\gamma\pi
N$ amplitude. Here again the diagrams that contribute to the sub-amplitude can
be disconnected provided the diagrams that are summed to give the full
amplitude are the sum of connected diagrams. The second class of diagrams are
those with no $\gamma\pi N$ intermediate states. The sum of these diagrams we
denote as $M_\pi^{(2)}$. The subscript $_\pi$ indicates that three-particle
intermediate states have only pions, i.e.\ $\pi\pi N$ intermediate states. Thus
the two-particle irreducible $\pi N\leftarrow\gamma N$ amplitude can be
written as
\begin{equation}
   M^{(2)}=M_\pi^{(2)}
      +\left\{\tilde{F}^{(2)}_{3}\,\tilde{G}^{(2)}\,
              \tilde{F}_{2}^{(3)\dagger}\right\}_{c} \ . \label{eq:3.25}
\end{equation}
At this stage we could parameterize $M_\pi^{(2)}$ as an energy independent
amplitude without affecting the two-body unitarity or the inclusion of the
$\gamma\pi N$ threshold of our final amplitude. However, if we want to relate
our input to the underlying Lagrangian given in Eq.~(\ref{eq:3.1}), we need to
examine the structure of $M_\pi^{(2)}$. In particular, we need to divide the
diagrams that contribute to this amplitude into two classes. Those with no
$\pi\pi N$ intermediate states, and therefore three-particle irreducible.
These we denote by $M^{(3)}$, and for the Lagrangian under consideration
reduce to the contribution of just the basic term in the Lagrangian, ${\cal
L}_{\gamma\pi NN}$. The second class of diagrams has $\pi\pi N$ intermediate
states, and need to be included if three-particle unitarity is to be included
in the final photoproduction amplitude. However, since we are neglecting
three-body unitarity we have a choice of either neglecting the contribution of
these diagrams to the $\pi N\leftarrow\gamma N$ amplitude $M_\pi^{(2)}$, or
examining the structure of these diagrams with the help of the last-cut lemma
in conjunction with the modified version of Taylor's classification of
diagrams~\cite{PA94}. Here we should point out that the contribution to
$M_\pi^{(2)}$ from diagrams which admit a $(\pi\pi N)$-cut that cut initial,
final and internal lines will be included  in $\left\{\tilde{F}^{(2)}_{3}\,
\tilde{G}^{(2)}\,\tilde{F}_{2}^{(3)\dagger}\right\}_{c}$. That leaves diagrams
in which the $(\pi\pi N)$-cut can intersect initial and internal lines only,
or final and internal lines only. Because we are including the e.m.\
interaction to first order, and have excluded any direct coupling between a
$\pi\pi\pi N$ intermediate state and the $\gamma N$ initial state, there are
no diagrams in this class. This leaves the diagrams that admit $(\pi\pi
N)$-cuts that intersect initial and internal lines only. These will involve
the connected $\pi N\leftarrow\pi\pi N$ diagrams which have the contribution
of three-body unitarity. For the present investigation we have chosen to
neglect this contribution to three-particle unitarity and have taken
$M_\pi^{(2)} = M^{(3)}$, and therefore
\begin{equation}
M^{(2)} = M^{(3)} +
          \left\{ \tilde{F}^{(3)}_{3}\,\tilde{G}^{(2)}\,
          \tilde{F}_{2}^{(2)\dagger}\right\}_{c}\ .      \label{eq:3.26}
\end{equation}

We now turn our attention to the contribution from the $\gamma\pi N$ threshold
to
$M^{(2)}$, i.e.\ $\left\{\tilde{F}^{(3)}_{3}\, \tilde{G}^{(2)}\,
\tilde{F}_{2}^{(2)\dagger}\right\}_{c}$. Since the e.m.\ coupling is included
to first order in the present analysis, we have that the $\gamma\pi N
\leftarrow \gamma N$ amplitude $\tilde{F}_2^{(2)\dagger}$ has to be
disconnected and  of the form
\begin{equation}
 \tilde{F}_{2}^{(2)\dagger}=\tilde{F}_{2;d}^{(2)\dagger}
                         =D^{-1}\Lambda^{(1)\dagger}_5 \ .\label{eq:3.27}
\end{equation}
On the other hand the three-particle  irreducible $\pi N\leftarrow\gamma\pi N$
amplitude $\tilde{F}^{(3)}_{3}$ can be written in terms of a connected and
disconnected part as
\begin{equation}
 \tilde{F}^{(3)}_{3}=\tilde{F}^{(3)}_{3;d}
                    +\tilde{F}^{(3)}_{3;c}\ .             \label{eq:3.28}
\end{equation}
Since the photon is absorbed in this process, the  disconnected amplitude
$\tilde{F}^{(3)}_{3;d}$ has the form
\begin{equation}
   \tilde{F}^{(3)}_{3;d}=S_{0}^{-1}\Gamma^{\pi(2)}
                        +\Delta^{-1}\Gamma^{(2)} \ ,      \label{eq:3.29}
\end{equation}
where $\Gamma^{\pi(2)}$ is the two-particle irreducible
$\pi\leftarrow\gamma\pi$ amplitude. Since we already have a spectator pion,
and to avoid the problem of non-linearity of our final integral equations, we
could take
\begin{equation}
\Gamma^{(2)}\rightarrow \Gamma^{(3)} = \Gamma\ ,         \label{eq:3.30}
\end{equation}
where $\Gamma$ is given in Eq.~(\ref{eq:2.17}). Since there is no direct
coupling in the Lagrangian between the final $\pi N$ state and any state
with four particles (e.g.\ $\pi\pi\pi N$ or $\gamma\pi\pi N$), we take
$\tilde{F}^{(3)}_{3;c}$ to be the $\pi N\leftarrow\gamma\pi N$ interaction in
the Lagrangian, i.e.\
\begin{equation}
\tilde{F}^{(3)}_{3;c} = \langle\pi N|-{\cal L}_{\pi N\gamma\pi N}
|\gamma\pi N\rangle\ .                                    \label{eq:3.31}
\end{equation}
As with the exposure of the cross diagram in Eq.~(\ref{eq:3.4}),
$\tilde{F}^{(3)}_{3;c}$ can also have the last-cut lemma further applied before
contact with the underlying Lagrangian is made.  At this stage this will be
ignored since we are only interested in the {\em minimal} requirements of
two-body unitarity and so $t^{(2)}$ will be considered as a background $\pi N$
potential.

We now make use of Eqs.~(\ref{eq:3.27}) and (\ref{eq:3.29}) in
Eq.~(\ref{eq:2.26})  to write the two-particle irreducible amplitude for $\pi
N\leftarrow\gamma N$ as
\begin{eqnarray}
   M^{(2)} &=& M^{(3)} + \tilde{F}^{(3)}_{3;c}\,g\,\Lambda^{(1)\dagger}_5
                       + \Gamma^{\pi(2)}\,\Delta\,\Lambda^{(1)\dagger}_5
                       + \Gamma^{(2)}\,S_0\,\Lambda^{(1)\dagger}_5\nonumber\\
           &=& M^{(3)} + \left[\tilde{F}^{(3)}_{3;c}
                         + \Gamma^{\pi(2)}\,S_0^{-1}
                         + \Gamma^{(2)}\Delta^{-1}\right]\,
                         g\,\Lambda^{(1)\dagger}_5 \ .      \label{eq:3.32}
\end{eqnarray}
This basically consists of the `seagull' term ($M^{(3)}$), the nucleon
emitting a pion with the photon being absorbed on the $\pi N$ interaction.
The details of this $\pi N\leftarrow\gamma\pi N$ amplitude will be determined
by the gauging of the $\pi N$ interaction. The two final terms on the right
hand side of Eq.~(\ref{eq:3.32}) correspond to photon absorption on the pion
and
the crossed diagrams for pion photoproduction.

We now can write the Born amplitude for pion photoproduction $\tilde{v}$
given in Eq.~(\ref{eq:3.18}) as
\begin{equation}
\tilde{v} = M^{(3)} + \left[\tilde{F}^{(3)}_{3;c}
                      + \Gamma^{\pi(2)}\,S_0^{-1}
                      + \Gamma^{(2)}\, \Delta^{-1}\right]\,g\,
                        \Lambda^{(1)\dagger}_5
             + \Lambda_5^{(2)\dagger}\,S_0\,\Gamma^{(2)}\ .\label{eq:3.33}
\end{equation}
For the present, the `seagull' term is the contact term in the Lagrangian,
i.e.\
\begin{equation}
M^{(3)} = \Gamma^{CT} \ ,                                \label{eq:3.34}
\end{equation}
while the two-particle irreducible pion current
\begin{equation}
\Gamma^{\pi(2)}= \Gamma^\pi                              \label{eq:3.35}
\end{equation}
is given in Eq.~(\ref{eq:2.25}). For the minimal requirement  of two-body
unitarity and the inclusion of the $\pi N$ and $\gamma\pi N$ thresholds, we
could replace the reducibility of the $N\leftrightarrow\gamma N$ amplitude
$\Gamma^{(2)}$ and the $N\leftrightarrow\pi N$ amplitude $\Lambda^{(2)}_5$ by
an amplitude that is of higher reducibility, and which is directly related to
the basic terms in our Lagrangian, depending on the diagrams these amplitudes
contribute to. We will come back to this point when we consider the gauge
invariance of our overall amplitude.

\section{The pion photoproduction amplitude}\label{Sec.4}

Having derived a set of coupled integral equations for the pion
photoproduction amplitude, we proceed in this section to derive an expression
for this amplitude with the help of the Ward-Takahashi
identities~\cite{Wa50,Ta57} as employed by Kazes~\cite{Ka59}. We then can
compare the resultant amplitude, which is gauge invariant, to the unitary
amplitude derived in the last section.

The starting point for deriving a gauge invariant pion  photoproduction
amplitude is the $\pi NN$ three-point function or Green's function for $\pi
N\leftarrow N$, given by
\begin{eqnarray}
   G(q,p',p) &=& \int d^4\!x_1d^4\!x_2d^4\!x_3
               e^{i(p'\cdot x_3 + q\cdot x_1 - p\cdot x_2)}
             \ \langle 0 | T\bigl[\phi(x_1)\psi(x_2)\bar{\psi}(x_3)
               \bigr]| 0 \rangle \nonumber \\
      &=&S(p')\,\Delta(q)\,\Lambda^{\dagger}_5(q,p',p)\,
          \vec{\tau}\cdot\vec{\phi}\,S(p) \ ,            \label{eq:4.1}
\end{eqnarray}
where $q$ is the pion momentum and $p'$ ($p$) is the momentum of the final
(initial) nucleon. Here $S$ and $\Delta$ are the nucleon and pion propagators
respectively. In Eq.~(\ref{eq:4.1}) we have taken $\vec{\tau}$ to be the Pauli
isospin matrix and $\vec{\phi}$ to be the pion field. To gauge this amplitude
and thus generate a gauge invariant pion photoproduction amplitude, we make
use of the procedure of coupling the photon to all propagators and vertices.
This is equivalent to the substitutions~\cite{Ka59}
\begin{mathletters}
\label{eq:4.2}
\begin{eqnarray}
   S(p)&\rightarrow& S(p) + S(p')\,\Gamma_{\mu}(k,p',p)\,
                                        S(p)\,A^\mu     \label{eq:4.2a}\\
   \Delta(q)&\rightarrow& \Delta(q) +
              \Delta(q')\,\Gamma^{\pi}_{\mu}(k,q',q) \,
                                      \Delta(q)\,A^\mu  \label{eq:4.2b}\\
   \Lambda^{\dagger}_5(q,p',p)\tau_i&\rightarrow&
   \Lambda^{\dagger}_5(q,p',p)\,\tau_i+
    \Gamma_{\mu}^{CTi}(k,q,p',p)\,A^{\mu}\ ,            \label{eq:4.2c}
\end{eqnarray}
\end{mathletters}
in the Green's function $G(q,p',p)$. To maintain unitarity we need to include
the pionic dressing of both the propagators and vertices. In particular, we
will find that the minimal requirement of two-body unitarity will impose a
constraint on the level of dressing in both the propagator $S(p)$ and
the vertex $\Lambda_5^{\dagger}(q,p',p)$.

With the above gauging procedure, our $\pi NN$ Green's function can be
gauged to give
\begin{equation}
   G(q,p',p) \rightarrow G(q,p',p)
  + S(p')\,\Delta(q)\,M_{\mu;\pi N\leftarrow\gamma N}(q,k,p',p)\,
    S(p)\,A^\mu      \ ,                                \label{eq:4.3}
\end{equation}
The resultant pion photoproduction  amplitude $M_{\mu;\pi N\leftarrow\gamma
N}(q,k,p',p)$ is now given as the sum of four diagrams, i.e.\
\begin{eqnarray}
   M^{i}_{\mu;\pi N\leftarrow\gamma N}(q,k,p',p)
   &=&\Lambda^{\dagger}_5(q,p',p+k)\,\tau_i\,S(p+k)\,
              \Gamma_{\mu}(k,p+k,p)                      \nonumber \\
    &&\quad + \Gamma_{\mu}(k,p',p'-k)\,S(p'-k)\,
              \Lambda^{\dagger}_5(q,p'-k,p)\,\tau_{i} \nonumber \\
   &&\quad +\Gamma^{\pi ij}_{\mu}(k,q,q-k)\,\Delta(q-k)\,
              \Lambda^{\dagger}_5(q-k,p',p)\,\tau_{j} \nonumber \\
   &&\quad +\Gamma_{\mu}^{CTi}(k,q,p',p)\ .             \label{eq:4.4}
\end{eqnarray}
Here $q$ is the pion momentum, $k$ the photon momentum  while $p'$ and $p$ are
the nucleon momenta in the final and initial states respectively.  The
superscripts $i$ and $j$ in $M^{i}_{\mu;\pi N\leftarrow\gamma N}$ and
$\Gamma^{\pi ij}_{\mu}$ indicate the isospin index, which we have now
explicitly included. At this stage the pionic dressing in the $\pi
N\leftarrow N$ amplitude $\Lambda^{\dagger}_5$, the nucleon current
$\Gamma_\mu$, the pion current $\Gamma^\pi_\mu$ and `seagull' term
$\Gamma^{CT}_\mu$ are included to all orders. These four diagrams are
illustrated in Fig.~\ref{Fig.7} and at this stage give the most general form
for the $\pi N\leftarrow\gamma N$ amplitude if we ignore the irreducibility
of the sub-amplitudes in the figure~\cite{Ka59,NK90,GG54}.

Using the Ward-Takahashi identities~\cite{Wa50,Ta57} for the nucleon
propagator, the pion propagator and the $\pi NN$ vertex, we can establish the
gauge invariance of the amplitude $M^{i}_{\mu;\pi N\leftarrow\gamma N}$. In
particular we have that $k^{\mu}M^{i}_{\mu;\pi N\leftarrow\gamma N}$ reduces
to
\begin{eqnarray}
 k^{\mu}M^{i}_{\mu;\pi N\leftarrow\gamma N}(q,k,p',p)
        &=&e_{N}\,\tau_{i}\,S^{-1}(p')\,S(p'-k)\,
                 \Lambda^{\dagger}_5(q,p'-k,p) \nonumber \\
         &&\quad -\tau_{i}\,e_{N}\,\Lambda^{\dagger}_5(q,p',p+k)
                  \,S(p+k)\,S^{-1}(p)             \nonumber \\
         &&\quad -ie\,\epsilon_{3ij}\,\tau_{j}\,\Delta^{-1}(q)\,\Delta(q-k)\,
                  \Lambda^{\dagger}_5(q-k,p',p)\ ,     \label{eq:4.5}
\end{eqnarray}
which is in agreement with the results of Kazes~\cite{Ka59}. Taking  matrix
elements of equation~(\ref{eq:4.5}) between on-mass shell initial and final
states gives
\begin{equation}
   \langle \pi N|
   k^{\mu}M^{i}_{\mu;\pi N\leftarrow\gamma N}(q,k,p',p)|\gamma N
   \rangle = 0\ ,                                       \label{eq:4.6}
\end{equation}
since the inverse propagators are zero on-mass shell. In this way we  have
established the fact that $M^{i}_{\mu;\pi N\leftarrow\gamma N}$ is indeed
gauge invariant and the requirement of current conservation is satisfied.

At this stage the $\pi NN$ Green's function we have considered has the fully
dressed propagators and vertex, and to get these quantities will require a
full solution to the underlying field theory. However, to satisfy the minimum
requirement of two-body unitarity, we need not include the pionic dressing of
the final nucleon propagator, while the initial nucleon propagator and the
$\pi N\leftarrow N$ amplitude need only include the minimal pionic dressing to
include the $\pi N$ threshold. In this case the Green's function we need to
consider is given by
\begin{equation}
   G_{1}(q,p',p) = S_{0}(p')\,\Delta(q)\,\Lambda^{(1)\dagger}_5(q,p',p)\,
               \vec{\tau}\cdot\vec{\phi}\,S^{(1)}(p) \ ,  \label{eq:4.7}
\end{equation}
then this Green's function can be gauged to give a corresponding amplitude
for pion photoproduction that satisfies the {\em minimum} requirement of
two-body unitarity.  Here the final nucleon propagator is gauged by the
transformation given in Eq.~(\ref{eq:2.19}), while the gauging of the $\pi
N\leftarrow N$ amplitude $\Lambda^{(1)\dagger}_5$ and the dressed nucleon
propagator
$S^{(1)}$ will be derived using the results of the previous section. The
result of gauging the Green's function  in Eq.~(\ref{eq:4.7}) is
\begin{equation}
   G_{1}(q,p',p)\rightarrow G_{1}(q,p',p) + S_{0}(p')\,\Delta(q)
              \,\overline{M}_{\mu;\pi N\leftarrow\gamma N}(q,k,p',p)\,
                 S^{(1)}(p)\,A_\mu       \ ,                \label{eq:4.8}
\end{equation}
where the $\pi N\leftarrow\gamma N$ amplitude $\overline{M}^{\,i}_{\mu;\pi
N\leftarrow\gamma N}(q,k,p',p)$ is given by
\begin{eqnarray}
 \overline{M}^{\,i}_{\mu;\pi N\leftarrow\gamma N}(q,k,p',p)
   &=&\Lambda^{(1)\dagger}_5(q,p',p+k)\,\tau_{i}\,S^{(1)}(p+k)\,
      \Gamma^{(1)}_{\mu}(k,p+k,p)                     \nonumber \\
   &&\quad  + \Gamma^{0}_{\mu}(k,p',p'-k)\,S_{0}(p'-k)\,
          \Lambda^{(1)\dagger}_5(q,p'-k,p)\,\tau_{i}  \nonumber \\
   &&\quad  + \Gamma^{\pi ij}_{\mu}(k,q,q-k)\,\Delta(q-k)\,
          \Lambda^{(1)\dagger}_5(q-k,p',p)\,\tau_{j}  \nonumber \\
   &&\quad  + \Gamma_{\mu}^{CT(1)i}(k,q,p',p)\ .          \label{eq:4.9}
\end{eqnarray}
This result differs from that of Eq.~(\ref{eq:4.4})  to the extent that the
nucleon in the final state did not require pionic dressing for two-body
unitarity to be satisfied, and as a result the corresponding current,
$\Gamma_\mu^0$ does not include any pionic corrections. On the other hand the
current resulting from gauging the initial nucleon has the necessary pionic
corrections to satisfy two-body unitarity. The proof that this amplitude
satisfies gauge invariance follows the same procedure considered above, with
the difference being the difference in the Ward-Takahashi identities for the
bare and dressed nucleon propagators $S_0$ and $S$. In this case the gauge
invariance of the amplitude
$\overline{M}^{\,i}_{\mu;\pi N\leftarrow\gamma N}$ is given by
\begin{eqnarray}
   k^{\mu}\overline{M}^{\,i}_{\mu;\pi N\leftarrow\gamma N}(q,k,p',p)
   &=&e_{N}\,\tau_{i}\,S^{-1}_{0}(p')\,S_{0}(p'-k)\,
      \Lambda^{(1)\dagger}_5(q,p'-k,p)               \nonumber \\
   &&\quad -\tau_{i}\,e_{N}\,\Lambda^{(1)\dagger}_5(q,p',p+k)\,
       S^{(1)}(p+k)\,S^{(1)-1}(p)                    \nonumber \\
   &&\quad -i\,e\,\epsilon_{3ij}\,\tau_{j}\,\Delta^{-1}(q)\,\Delta(q-k)\,
        \Lambda^{(1)\dagger}_5(q-k,p',p)\ ,             \label{eq:4.10}
\end{eqnarray}
which on-mass shell gives a conserved current, provided both the bare and
dressed nucleon propagators have poles at the physical nucleon mass. Thus we
have constructed a gauge invariant amplitude for pion photoproduction starting
with the Green's function for the process $\pi N\leftarrow N$ in which the
final nucleon propagator has less pionic dressing than the propagator in the
initial state. We proceed in the next section to demonstrate that this gauge
invariant amplitude is identical to that resulting from the solution of the
coupled equations, and that it satisfies two-body unitarity.

\section{Is the Gauge Invariant Amplitude Unitary?}\label{Sec.5}

In the previous section we demonstrated how one may construct a gauge
invariant amplitude for pion photoproduction given the $\pi N\leftarrow N$
Green's function and the transformation of the pion propagator, nucleon
propagator and the $\pi N\leftarrow N$ amplitude under the gauge symmetry. In
Sec.~\ref{Sec.2} these gauge transformations were derived for the different
terms in the basic meson-baryon Lagrangian. Here we are going to use the
gauging
of these basic terms in the Lagrangian to derive the results of gauging the
pionic dressed nucleon propagator $S(p)$, the pion propagator $\Delta(q)$ and
the one-particle irreducible $\pi N\leftarrow N$ amplitude
$\Lambda_5^{(1)\dagger}$. From this point on we will drop the superscript
$^{(1)}$ on the dressed nucleon propagator assuming that $S(p)$ includes the
 necessary pionic dressing to include two-body unitarity. These results when
used in the definition of the three-point function will give us the amplitude
for pion photoproduction with the application of LSZ reduction to the resultant
Green's function. The amplitude resulting from this gauging procedure is
identical to that resulting from the solution of the coupled set of integral
equations derived in Sec.~\ref{Sec.3}. In this way we establish the gauge
invariance of the pion photoproduction amplitude resulting from the solution
of the coupled integral equations, which already satisfy unitarity.

Since the nucleon propagator $S_0(p')$ need not include pionic dressing, but
might include the quark-gluon structure, we may use the result of
Sec.~\ref{Sec.2} to write
\begin{equation}
   S_0\stackrel{\mbox{U(1)}}{\rightarrow}S_0
      +S_0\,\Gamma^{(3)}\,S_0\ .                         \label{eq:5.1}
\end{equation}
If we assume that our `bare' nucleon has the quark-gluon structure, and this
structure does not contribute to unitarity, then the nucleon current is
$\Gamma^{(3)}_\mu=\Gamma_\mu$ with $\Gamma_\mu$ given in Eq.~(\ref{eq:2.17}).
Note that in the absence of the quark-gluon structure, this current will
reduce to the standard Dirac current for a point Fermion.  In a similar
manner, the gauge prescription for the basic input into our coupled integral
equations, which are also the terms in the Lagrangian, are given as
\begin{eqnarray}
   \Lambda_5^{(2)\dagger}\stackrel{\mbox{U(1)}}{\rightarrow}
   \Lambda_5^{(2)\dagger}+\Gamma^{CT(3)}                  \label{eq:5.2}\\
   \Lambda_5^{(2)}\stackrel{\mbox{U(1)}}{\rightarrow}
   \Lambda_5^{(2)}+\Gamma^{(3)}_\pi                       \label{eq:5.3}\\
   t^{(2)}\stackrel{\mbox{U(1)}}{\rightarrow}t^{(2)}
   +\tilde{F}_{3;c}^{(3)} \ .                             \label{eq:5.4}
\end{eqnarray}
Here, in addition to the gauging of the $\pi NN$ vertex, we have included the
gauging of the two-particle irreducible $\pi N\leftarrow \pi N$ amplitude. For
the pion propagator we take
\begin{equation}
   \Delta\stackrel{\mbox{U(1)}}{\rightarrow}\Delta+
\Delta\,\Gamma^{\pi(1)}\,\Delta \ .                       \label{eq:5.5}
\end{equation}
In our meson-baryon Lagrangian the only dressing the pion can have is via
nucleon antinucleon loops~\cite{Foot1}. If we ignore these loop corrections to
the pion dressing, the only dressing we could have are the result of the
underlying quark-gluon structure. The pion current in this case can be written
in terms of the gauge invariant current given in Eq.~(\ref{eq:2.25}), i.e.\
$\Gamma^{\pi(1)}_\mu = \Gamma^{\pi(2)}_\mu = \Gamma^{\pi}_\mu$.

To gauge the one-particle irreducible $\pi N$ amplitude $t^{(1)}$, we need to
first gauge the $\pi N$ propagator $g$. Since only two-body unitarity need be
considered, the nucleon propagator in $g$ can be taken to be the bare
propagator $S_0$, and the $\pi N$ propagator reduces to $g=S_0\Delta$. To first
order in the e.m.\ coupling the gauge transformation of the $\pi N$ propagator
reduces to
\begin{eqnarray}
   g&\stackrel{\mbox{U(1)}}{\rightarrow}& g
            + \Delta\, S_0\,\Gamma^{(3)}\,S_0
            + S_0\,\Delta\,\Gamma^{\pi(2)}\Delta \nonumber \\
        &=& g + g\,\tilde{F}^{(3)}_{3;d}\,g\ .          \label{eq:5.6}
\end{eqnarray}
With the help of the two-body equation for $t^{(1)}$, and the transformation
properties of the $\pi N$ amplitude $t^{(2)}$ and $g$, the gauging of the one
particle irreducible $\pi N$ amplitude $t^{(1)}$ reduces to
\begin{equation}
   t^{(1)}\stackrel{\mbox{U(1)}}{\rightarrow}t^{(1)}
                                +\tilde{F}_3\ ,         \label{eq:5.7}
\end{equation}
where to first order in the e.m.\ coupling
\begin{equation}
 \tilde{F}_3 = \left(t^{(1)}g+1\right)\tilde{F}_{3;c}^{(3)}
                          \left(1+gt^{(1)}\right)
             +t^{(1)}\,g\,\tilde{F}^{(3)}_{3;d}\,g\,t^{(1)}\ .\label{eq:5.8}
\end{equation}
With these results in hand, we can now determine how $\Lambda_5^{(1)\dagger}$
behaves under gauging. Given that the one-particle irreducible amplitude for
$\pi N\leftarrow N$ is given, using Eq.~(\ref{eq:3.5}), by the relation
\begin{equation}
\Lambda^{(1)\dagger}_5 = \left(t^{(1)}\,g+1\right)\,
                            \Lambda^{(2)\dagger}_5\ ,       \label{eq:5.9}
\end{equation}
we can make use of Eqs.~(\ref{eq:5.2}), (\ref{eq:5.6}) and (\ref{eq:5.7}) to
write the gauge transformation for $\Lambda^{(1)\dagger}_5$ as
\begin{equation}
   \Lambda_5^{(1)\dagger}\stackrel{\mbox{U(1)}}{\rightarrow}
               \Lambda_5^{(1)\dagger} + M' \ ,       \label{eq:5.10}
\end{equation}
where to first order in e.m.\ coupling the $\pi N\leftarrow\gamma N$ amplitude
$M'$ is given by
\begin{equation}
   M' = \left(t^{(1)}\,g+1\right)\left(\Gamma^{CT}
              + \tilde{F}^{(3)}_{3;c}\,g\,\Lambda^{(1)\dagger}_5\right)
      + t^{(1)}\,g\,\tilde{F}^{(3)}_{3;d}\,g\,\Lambda^{(1)\dagger}_5
                                                   \ .\label{eq:5.11}
\end{equation}
All that is left to determine is how the nucleon propagator $S$ behaves under
gauging when the $\pi N$ unitarity cut has been exposed. From
Eqs.~(\ref{eq:3.6}) and (\ref{eq:3.7}) we have that
\begin{eqnarray}
  S &=& \left[ S_0^{-1} - \Sigma^{(1)}\right]^{-1}\nonumber \\
    &=& S_0 + S_0\,\Sigma^{(1)}\,S\ ,                 \label{eq:5.12}
\end{eqnarray}
where
\begin{equation}
\Sigma^{(1)} = \Lambda^{(2)}_5\,g\,\Lambda^{(1)\dagger}_5\ .\label{eq:5.13}
\end{equation}
With the help of Eqs.~(\ref{eq:5.1}), (\ref{eq:5.3}), (\ref{eq:5.6}) and
(\ref{eq:5.10}), the dressed nucleon propagator can be gauged to give
\begin{equation}
   S^{-1}\stackrel{\mbox{U(1)}}{\rightarrow}S^{-1}
             -\tilde{\Gamma}^{(1)} \ ,                        \label{eq:5.14}
\end{equation}
where the dressed $N\leftarrow\gamma N$ amplitude $\tilde{\Gamma}^{(1)}$ is
given, to first order in the e.m.\ coupling, by
\begin{eqnarray}
 \tilde{\Gamma}^{(1)}&=& \Gamma^{(3)}
                  + \Gamma_\pi^{(3)}\,g\,\Lambda^{(1)\dagger}_5
                  + \Lambda^{(2)}_5\,g\,\left[ \tilde{F}^{(3)}_{3;d}\,g\,
                    \Lambda^{(1)\dagger}_5 + M'\right]\nonumber \\
         &=& \Gamma^{(2)} + \Lambda^{(1)}\,g\,\left[ \Gamma^{CT}
            + \tilde{F}^{(3)}_3\,g\,\Lambda^{(1)\dagger}_5\right]
                                    \ .                       \label{eq:5.15}
\end{eqnarray}
In writing the second line in Eq.~(\ref{eq:5.15}), we have made use of
Eqs.~(\ref{eq:3.22}) and (\ref{eq:5.11}). If we now compare the dressed
nucleon current resulting from the gauging of the dressed nucleon propagator,
and make use of Eqs.~(\ref{eq:3.16}) and (\ref{eq:3.32}), we find that this
current is in fact identical to the one-particle irreducible current
derived in Sec.~\ref{Sec.3}, i.e.\
\begin{equation}
   \tilde{\Gamma}^{(1)}=\Gamma^{(1)} \ .                      \label{eq:5.16}
\end{equation}
Therefore the nucleon propagator $S$ behaves, to first order in the
electromagnetic coupling, under gauging as
\begin{equation}
   S\stackrel{\mbox{U(1)}}{\rightarrow}S
          +S\,\Gamma^{(1)}\,S \ .                        \label{eq:5.17}
\end{equation}
Thus exposing the $\pi N$ unitarity cut in $S$, and gauging, leads to the same
results as exposing the $\gamma\pi N$ unitarity cut which is what one expects.

To complete the proof of the equivalence of the two formulations of
Secs.~\ref{Sec.3} and \ref{Sec.4} we are required to show that
\begin{equation}
   M=M^{(0)}\ ,                                           \label{eq:5.18}
\end{equation}
where $M$ results from substituting Eqs.~(\ref{eq:5.17}), (\ref{eq:5.10}),
(\ref{eq:5.5}) and (\ref{eq:5.1}) into Eq.~(\ref{eq:4.7}) with the result
that gauging the three-point function for the process $\pi N\leftarrow N$ is
given by
\begin{equation}
   G_1\stackrel{\mbox{U(1)}}{\rightarrow}G_1
        +S_0\,\Delta\ M\ S                                \label{eq:5.19}
\end{equation}
and the corresponding pion photoproduction amplitude $M$ is of the form
\begin{equation}
   M=\Gamma^{(3)}\,S_0\,\Lambda_5^{(1)\dagger}
    +\Gamma^{\pi(1)}\,\Delta\,\Lambda_5^{(1)\dagger}
    +\Lambda_5^{(1)\dagger}\,S\,\Gamma^{(1)}+M'  \ .      \label{eq:5.20}
\end{equation}
In writing Eq.~(\ref{eq:5.19}), we have restricted ourselves to first order in
the electromagnetic coupling.

Substituting for $M'$ from  Eq.~(\ref{eq:5.11}) and making use of the
definition of the two-particle irreducible $\pi N\leftarrow\gamma N$ amplitude
$M^{(2)}$ given in Eq.~(\ref{eq:3.32}), we get
\begin{eqnarray}
M &=& \left(t^{(1)}\,g+1\right)\,M^{(2)}
      + \Lambda^{(1)\dagger}_5\,S\,\Gamma^{(1)}  \nonumber \\
  &=& M^{(2)} + t^{(2)}\,g\,M^{(1)}
      + \Lambda^{(1)\dagger}_5\,S\,\Gamma^{(1)}\ .  \label{eq:5.21}
\end{eqnarray}
In writing the second line we have made use of Eq.~(\ref{eq:3.15}) to relate
the one-particle and two-particle irreducible $\pi N\leftarrow\gamma N$
amplitudes. Making use of the adjoint of Eq.~(\ref{eq:5.3})
we can write $M$ as
\begin{eqnarray}
M &=& M^{(2)} + \Lambda^{(2)\dagger}_5\,S\,\Gamma^{(1)}
     + t^{(2)}\,g\,\left[M^{(1)}
      + \Lambda^{(1)\dagger}_5\,S\,\Gamma^{(1)}\right] \nonumber \\
  &=& M^{(2)} + \Lambda^{(2)\dagger}_5\,S_0
          \left[\Gamma^{(2)} + \Lambda^{(2)}_5\,g\,M^{(0)}\right]
          + t^{(2)}\,g\,M^{(0)}\ .                    \label{eq:5.22}
\end{eqnarray}
The second line of Eq.~(\ref{eq:5.22}) is derived by making use of
Eq.~(\ref{eq:3.19}) to write $S\,\Gamma^{(1)}$ in terms of $M^{(0)}$, and
Eq.~(\ref{eq:3.14}) to get the last term on the right hand side. We now employ
the definition of the Born amplitude for pion photoproduction $\tilde{v}$,
Eq.~(\ref{eq:3.33}), and the $\pi N$ potential $v$ given in Eq.~(\ref{eq:3.12})
to write Eq.~(\ref{eq:5.22}) as
\begin{equation}
M = \tilde{v} + v\,g\,M^{(0)}  \ .                   \label{eq:5.23}
\end{equation}
If we compare this result with Eq.~(\ref{eq:3.17}), we observe that the right
hand side of both equations are identical. This proves the fact that the
amplitude which is a solution of the coupled equations that include the
$\gamma N$, $\pi N$ and $\gamma\pi N$ thresholds is identical to that derived
using the gauging of the $\pi N\leftarrow N$ Green's function with the help of
the Ward-Takahashi identities. Thus we have that
\begin{equation}
M = M^{(0)}\ ,                                      \label{eq:5.24}
\end{equation}
and in this way we have established the fact that the solution of the coupled
integral equation, Eq.~(\ref{eq:3.17}), gives us an amplitude that satisfies
two-body unitarity, and is gauge invariant.

To establish the connection with previous results in the literature, we
iterate Eq.~(\ref{eq:5.23}) with the result
\begin{eqnarray}
M^{(0)} &=& \tilde{v} + \left(v+v\,g\,v + v\,g\,v\,g\,v + \cdots\right)\,
             \tilde{v}   \nonumber \\
        &=& \left(t^{(0)}\,g+1\right)\,\tilde{v} \nonumber \\
        &=& \left(t^{(0)}\,g+1\right)\tilde{v}_B
            + \left(t^{(0)}\,g+1\right)\tilde{v}_R\ ,  \label{eq:5.25}
\end{eqnarray}
where $\tilde{v}_B$ is the Born term illustrated in Fig.~\ref{Fig.1} and has
often been used as a gauge invariant term on its own. The factor of
$\left(t^{(0)}\,g+1\right)$ gives the distortion in the $\pi N$ channel. The
second term, illustrated in Fig.~\ref{Fig.8}, is required to maintain gauge
invariance at the operator level for $M^{(0)}$, and results from including the
$\gamma\pi N$ channel into our integral equations. This inclusion of the
$\gamma\pi N$ channel effectively allows us to couple the photon not only to
the initial nucleon in the $\pi N\leftarrow N$ vertex, but also to the nucleon
in the final state. In this way we have coupled the nucleon to every
propagator and vertex in the three-point Green's function for $\pi N\leftarrow
N$.

\section{Conclusion}\label{Sec.6}

To employ pion photoproduction at medium energies to examine nucleon structure
and in particular the resonances observed in $\pi N$ scattering, we need a
theory that incorporates charged current conservation and the conservation of
probability, i.e.\ gauge invariance and unitarity. If in addition we would like
to test current models of QCD based on quark-gluon degrees of freedom, we need
to include form factors in which we have consistency between the $\pi NN$ and
$\gamma NN$ form factors resulting from QCD. In the present investigation, we
considered a Lagrangian written in terms of the meson-baryon degrees of
freedom. This Lagrangian has form factors which in principle could be derived
from a QCD model. We then gauged this Lagrangian with the help of the
procedure proposed by Ohta~\cite{Oh89,Oh90}. This gauging procedure could have
been circumvented had we considered a specific QCD model. However, to avoid
the problem of working within a specific model, we chose to maintain
generality and restrict our gauging to the approach proposed by Ohta.

Within the framework of this Lagrangian which includes not only the
meson-baryon degrees of freedom but also the coupling of the photon to these
degrees of freedom, we derived a set of coupled integral equations that
satisfy two-particle unitarity. This was achieved by exposing the $\pi N$,
$\gamma N$ and $\pi\gamma N$ thresholds following the analysis of
Taylor~\cite{Ta63,Ta66}. To prove that the solution of these coupled equations
is also gauge invariant, we employed the Ward-Takahashi identities,
derived for the basic terms in the Lagrangian, to gauge the one particle
irreducible three point function for the process $\pi N\leftarrow N$. Since we
demanded that our amplitude satisfy {\em only} two-body unitarity, we assumed
that the final nucleon propagator in the $\pi N\leftarrow N$ Green's function
did not have any pionic dressing, and the initial nucleon propagator had the
minimal dressing to include the $\pi N$ threshold, while the $\pi NN$ vertex in
the three-point function was one-particle irreducible. This gauging procedure
gave us an amplitude that is identical to that resulting from a solution of
coupled integral equations. We found it essential that we include the
$\gamma\pi N$ threshold into our coupled integral equation.

The need for the inclusion of the $\gamma\pi N$ threshold to satisfy
gauge invariance at the operator level, raises some questions regarding the
application of Watson's Theorem~\cite{Wa54} or its off-shell
modification~\cite{NB90} to unitarise the gauge invariant Born term
represented in Fig.~\ref{Fig.1}, and to maintain gauge invariance in the final
result. In fact when we cast our amplitude into a form of Watson's Theorem,
i.e.\ a distorted wave in the $\pi N$ channel, we found that we had an
additional term that has not been included in most calculations in the past.
These additional terms are required in order to preserve gauge invariance of
the photoproduction amplitude at the operator level. We are presently
considering a simple model to determine the contribution of these additional
terms required for maintaining gauge invariance and unitarity in the final
amplitude.

\acknowledgements

The authors would like to thank the Australian Research Grant Scheme for
their financial support and B.~C.~Pearce for some stimulating discussions.
One of the authors (IRA) would like to thank Professor J. Speth and the
hospitality of IKP at Forschungszentrum J\"ulich, where part of this work was
done.

\newpage\null

\begin{figure}[h]
\vskip 10 cm
\hskip 2 cm
\special{illustration Fig.1}
\caption{The Born amplitude for pion photoproduction. The number in the
vertices is the
irreducibility of the vertex.}\label{Fig.1}
\end{figure}

\vskip 0.5 cm

\begin{figure}[h]
\vskip 6 cm
\hskip 3 cm
\special{illustration Fig.2}
\vskip 0.5 cm
\caption{Two possible scenarios for constructing a gauge invariant Lagrangian
for
mesons and baryons from a QCD model}\label{Fig.2}
\end{figure}

\newpage\null

\begin{figure}[h]
\vskip 3 cm
\hskip 0.5 cm
\special{illustration Fig.3}
\vskip 0.5 cm
\caption{Examples of diagrams that could contribute to the baryon (a) and
meson (b) dressing in a quark model.}\label{Fig.3}
\end{figure}

\vskip 1 cm

\begin{figure}[h]
\vskip 4 cm
\hskip 4 cm
\special{illustration Fig.4}
\vskip 0.5 cm
\caption{The momentum labels for the $\pi NN$ vertex}\label{Fig.4}
\end{figure}

\vskip 1 cm

\begin{figure}[h]
\vskip 4 cm
\hskip 1 cm
\special{illustration Fig.5}
\vskip 0.5 cm
\caption{The Born amplitude for $\pi N$ scattering}\label{Fig.5}
\end{figure}

\newpage\null

\begin{figure}[h]
\vskip 8 cm
\hskip 1 cm
\special{illustration Fig.6 scaled 800}
\vskip 0.5 cm
\caption{The energy plane for $\pi + N \leftarrow \gamma + N$ scattering
showing the
unitarity cuts corresponding to the different thresholds}\label{Fig.6}
\end{figure}

\vskip 1 cm

\begin{figure}[h]
\vskip 6.5 cm
\hskip 3 cm
\special{illustration Fig.7 scaled 800}
\vskip 0.5 cm
\caption{The four types of diagrams that contribute to the pion
photoproduction amplitude. These result from the gauging of the $\pi NN$
three-point function.}\label{Fig.7}
\end{figure}

\newpage\null

\begin{figure}[h]
\vskip 7 cm
\hskip 0.5 cm
\special{illustration Fig.8 scaled 800}
\vskip 0.5 cm
\caption{The non-Born diagrams which contribute to the pion photoproduction
amplitude, the number in the circle gives the irreducibility of each
amplitude.}\label{Fig.8}
\end{figure}


\begin{thebibliography}{10}
\bibitem{Foot1}
If we include the $\sigma$ and $\rho$ exchange $\pi N$ potential into our basic
Lagrangian, then we have introduced  $\sigma\pi\pi$ and $\rho\pi\pi$
vertices into the theory and these in turn will contribute to pionic
dressing. For the present, since we have not specified the $\pi N$
interaction, we will ignore the dressing of the pion propagator.
\end{thebibliography}
\end{document}